\documentclass{iau}

\usepackage{amsmath}
\usepackage{graphicx}
\usepackage{multirow}

\begin{document}
% \lefttitle{Ananda Hota}
% \righttitle{RAD12: RAD@home citizen science discovery of an AGN spewing a large unipolar radio bubble onto its merging companion galaxy}

% \title[RAD12] %% give here short title %%
% {RAD12: RAD@home citizen science discovery of an AGN spewing a large unipolar radio bubble onto its merging companion galaxy}

\jnlPage{1}{7}
\jnlDoiYr{2023}
\doival{10.1017/xxxxx}

\aopheadtitle{Proceedings IAU Symposium 375}
\editors{abc, def. Valls-Gabaud, ghi.}

\title{RAD@home discovery of a one-sided radio jet hitting the companion galaxy} 
%RAD12: RAD@home citizen science discovery of an AGN spewing a large unipolar radio bubble onto its merging companion galaxy 
\author{Ananda Hota$^{1,2}$, Pratik Dabhade$^{3,2}$, Sravani Vaddi$^{4,2}$}
\affiliation{
$^1$UM-DAE Centre for Excellence in Basic Sciences, University of Mumbai, Mumbai-400098, India \\
$^2$RAD@home Astronomy Collaboratory, Kharghar, Navi Mumbai, 410210, India\\
$^3$Instituto de Astrof\' isica de Canarias, Calle V\' ia L\'actea, s/n, E-38205, La Laguna, Tenerife, Spain\\
$^4$Arecibo Observatory, NAIC, HC3 Box 53995, Arecibo, Puerto Rico, PR 00612, USA
}
\begin{abstract} 
Minkowski’s Object and `Death Star galaxy' are two of the famous cases of rare instances when a radio jet has been observed to directly hit a neighbouring galaxy. RAD12, the RAD@home citizen science discovery with GMRT being presented here, is not only a new system being added to nearly half a dozen rare cases known so far but also the first case where the neighbouring galaxy is not a minor/dwarf companion but a galaxy bigger than the host of the radio jet. Additionally, the jet appears to be one-sided and the jet after interaction completely stops and forms a bubble inflating laterally which is unlike previous cases of minor deviation or loss of collimation.  Since the nature of radio jet-ISM coupling is poorly understood so far, more discovery of objects like RAD12 will be important to the understanding of galaxy evolution through merger and AGN feedback.
\end{abstract}

\maketitle

% Galaxy: evolution, radio continuum: galaxies, methods: data analysis

\vspace{-0.5cm}
\section{Introduction}
Galaxy mergers and Active Galactic Nuclei(AGN) feedback have been the most favoured models to explain the observed properties of supermassive black holes and galaxies, namely the M-$\sigma$ relation, the colour-magnitude diagram, the galaxy luminosity function, etc. \citep{DiMatteo2005, HeckmanBest2014}.
Even after two decades of the proposed model, we are still in search of a ``smoking gun'' evidence showing the age of the AGN-driven outflow (cause) to precede the age of cessation of star formation (effect) in any sample of merger-remnant early-type galaxies \citep{Hota2016}. Furthermore, how the wind/jet fluid is coupled with the cold gas of the merger which may lead to momentarily positive or eventually negative feedback on star formation, is poorly understood. Here, we briefly report one extraordinary case, named RAD12, where radio jet outflow is seen interacting with the neighbouring galaxy in a possible major dry merger of two early-type galaxies. This was initially discovered by the RAD@home citizen science research collaboratory \citep{Hota2014} and then followed up with the upgraded Giant Meterwave Radio Telescope (GMRT).

The basic methodology of citizen science research by the RAD@home Collaboratory\footnote{\url{https://www.radathomeindia.org}} using GMRT telescope has been described in \citet{Hota2014,Hota2016}. Briefly, it can be stated as a nationwide Inter-University collaboratory of trained citizen scientists, professional astronomers and technical/administrative facilitators. It follows a hybrid method of both online and in-person training of citizens/students who have had basic University-level science education. The training is focused on ultraviolet(UV)-Optical-infrared(IR)-radio image analysis, through red-green-blue(RGB)-contour image overlays, by a large number of citizens/students who undergo training and report the discovery of interesting objects continuously. The platforms for interaction for growing the skill of RGB image interpretation are freely available Internet services, Facebook and Google. Citizens/students are trained in an organised way after a basic online selection process. Selected citizens who are trained extensively in a host institute for effectively one week are called e-astronomers. On the other hand, who are trained purely online or during one-day events with large gatherings of 50-200 members are called i-astronomers. One such one-day event was RAD@home Citizen Science Research Workshop organised during a science exhibition named {\it Vigyan Samagam} which was creating public awareness about the Square Kilometre Array (SKA). The week-long extensive training programmes are called {\it RAD@home Discovery Camps} which have been hosted by multiple national research institutes, research-cum-education institutes for undergraduate science and technology students, planetariums and science popularisation organisations of the Government of India.  After nearly ten years of successful demonstration international expansion of the Collaboratory has been kick-started by the education and outreach programme of the IAU symposium 375 (see Rajoria et al. in this Proceeding). During such training and later on from their respective homes, trained citizen scientists discover unusual radio sources which are then followed up with the GMRT through a GMRT Time Allocation Committee-approved project titled GMRT Observation of Objects Discovered by RAD@home Astronomy Collaboratory (GOOD-RAC). One such unusual object (RAD12) followed up with the GMRT is being presented in this paper focusing on jet-galaxy interaction. The general scientific motivation of the collaboratory is to discover spiral-host radio galaxies like Speca  \citep{Hota2011} and cases of AGN-jet feedback in the context of galaxy mergers like `cosmic leaf blower galaxy' NGC 3801 \citep{Hota2012}.

\vspace{-0.2cm}
\section{Discovery of RAD12}
\begin{figure}
   \centering
   \includegraphics[scale=0.2]{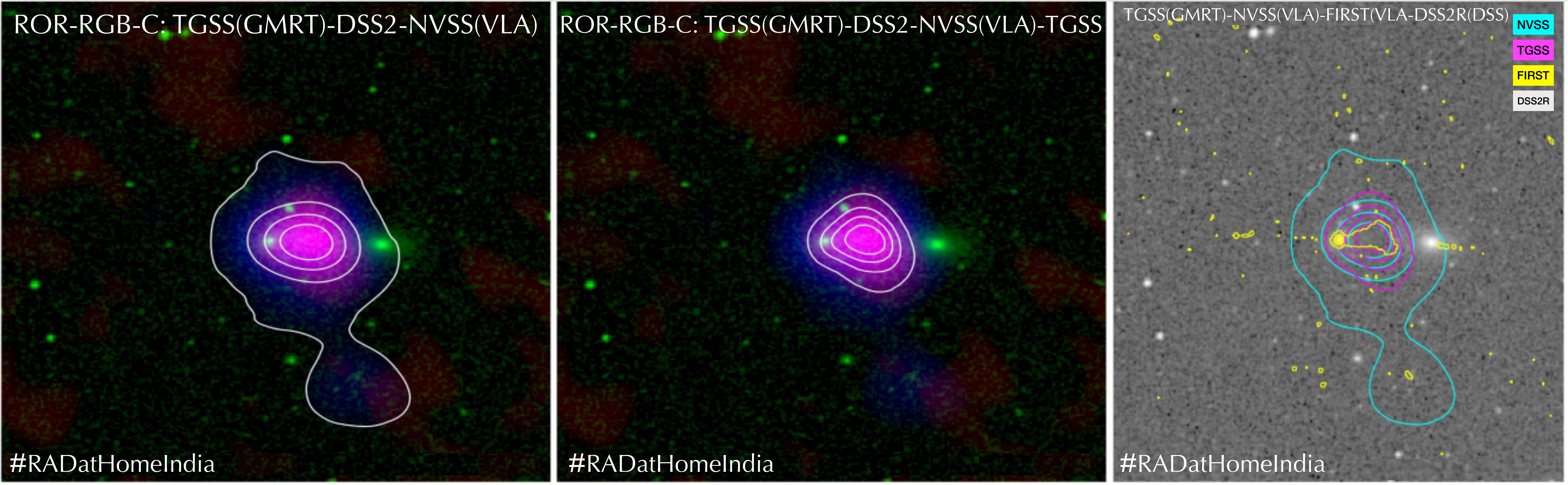}
      \caption{ Radio-Optical-Radio image overlay of RAD12 
in Red-Green-Blue channels and radio contours. This is a typical image output from RAD@home RGB-maker. The data on red, green, and blue channels are TIFR GMRT Sky Survey (TGSS ADR1), Digitized Sky Survey (DSS2 R) and NRAO VLA Sky Survey (NVSS) survey data. On the left panel, NVSS and on the middle TGSS radio contours are also plotted. The right panel plots the same optical data in greyscale along with TGSS (magenta), NVSS (cyan) and Faint Images of the Radio Sky at Twenty-Centimeters (FIRST; yellow) radio surveys. Note that the southern blob seen in NVSS is an unrelated radio source with optical and FIRST counterparts.}
   \label{fig:RGB}
\end{figure}
Radio-Optical-Radio Red-Green-Blue or ROR-RGB image overlays were the basic methods for trained citizen scientists to identify the host galaxy of a radio source. In the initial years, NASA SkyView was used but later we developed an RGB-maker web-tool (see Kumar et al. in this Proceeding). Figure.\ \ref{fig:RGB}, presents typical image outputs from the RGB-maker\footnote{\url{https://www.radathomeindia.org/rgbmaker}}. In such images, an FRII-type radio galaxy would have two pink-looking radio emission blobs on either side of a green-looking optical host galaxy. In the case of an FRI-type radio galaxy, the peak of the elongated radio emission should coincide with the optical host galaxy. However, a peculiar case was spotted where the pink-looking radio emission blob was located in between two optical galaxies, both of which are like ellipticals in our IR-optical-UV or IOU-RGB images. Further comparison of NVSS and FIRST contour images (Figure.\ \ref{fig:RGB}, right panel) clarified that the eastern optical galaxy coincides with the radio core, evidently the host, but NVSS emission, sensitive to diffuse emission, did not extend beyond the faint tail of emission seen in FIRST, sensitive to compact features. The faint radio tail seems to stop abruptly before reaching the western optical galaxy. It was suspected to be something else but not a head-tail radio galaxy, possibly a jet-galaxy interaction \citep{Hota2014}.

\begin{figure}
   \centering
   \includegraphics[scale=0.13]{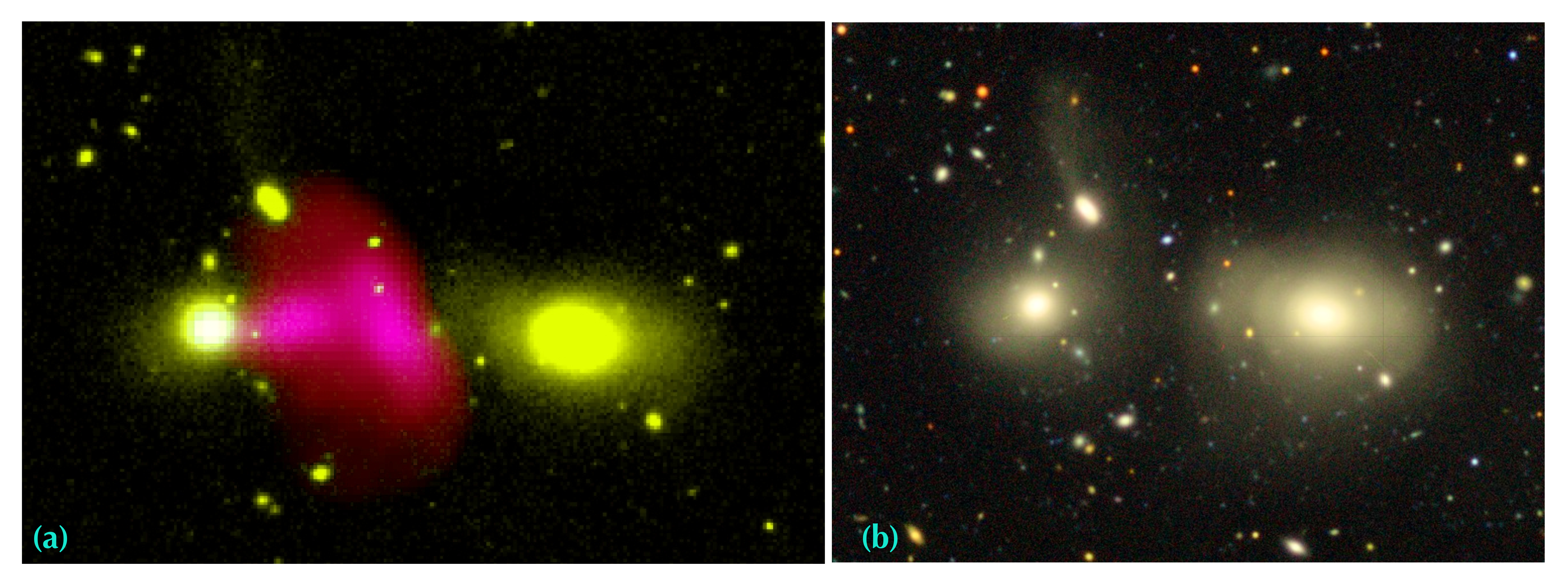}
   \caption{(a) Radio emission from GMRT in Red and MeerKAT in blue are overlaid on an optical image from Canada France Hawaii Telescope. Mushroom-shaped radio bubble seen in between two merging elliptical galaxies is clearly seen. (b) Deep optical image of RAD12 from legacy surveys, where the shells are prominently visible.}
   \label{fig:color}
\end{figure}

Details of the follow-up observations with the upgraded GMRT and of the supplementary archival data from the MeerKAT telescope are available in \citet{HOTA2022}. A collage of radio emission (GMRT 323 MHz in red and MeerKAT L-band in blue) along with CFHT r-band optical data, in yellow, has been presented in Figure.\ \ref{fig:color}a. GMRT detected the core-plume structure of the FIRST and resolved the NVSS blob to be a mushroom-shaped radio bubble. The tail of radio emission does not extend further to the west but shows a termination peak, supporting obstruction by the companion galaxy.  The accompanying deep optical image (Figure.\ \ref{fig:color}b) from the legacy\footnote{\url{https://www.legacysurvey.org/acknowledgment/}} survey clearly shows multiple stellar shells in the western companion galaxy. 

\vspace{-0.2cm}
\section{Discussion}

\begin{figure}
   \centering
   \includegraphics[scale=0.27]{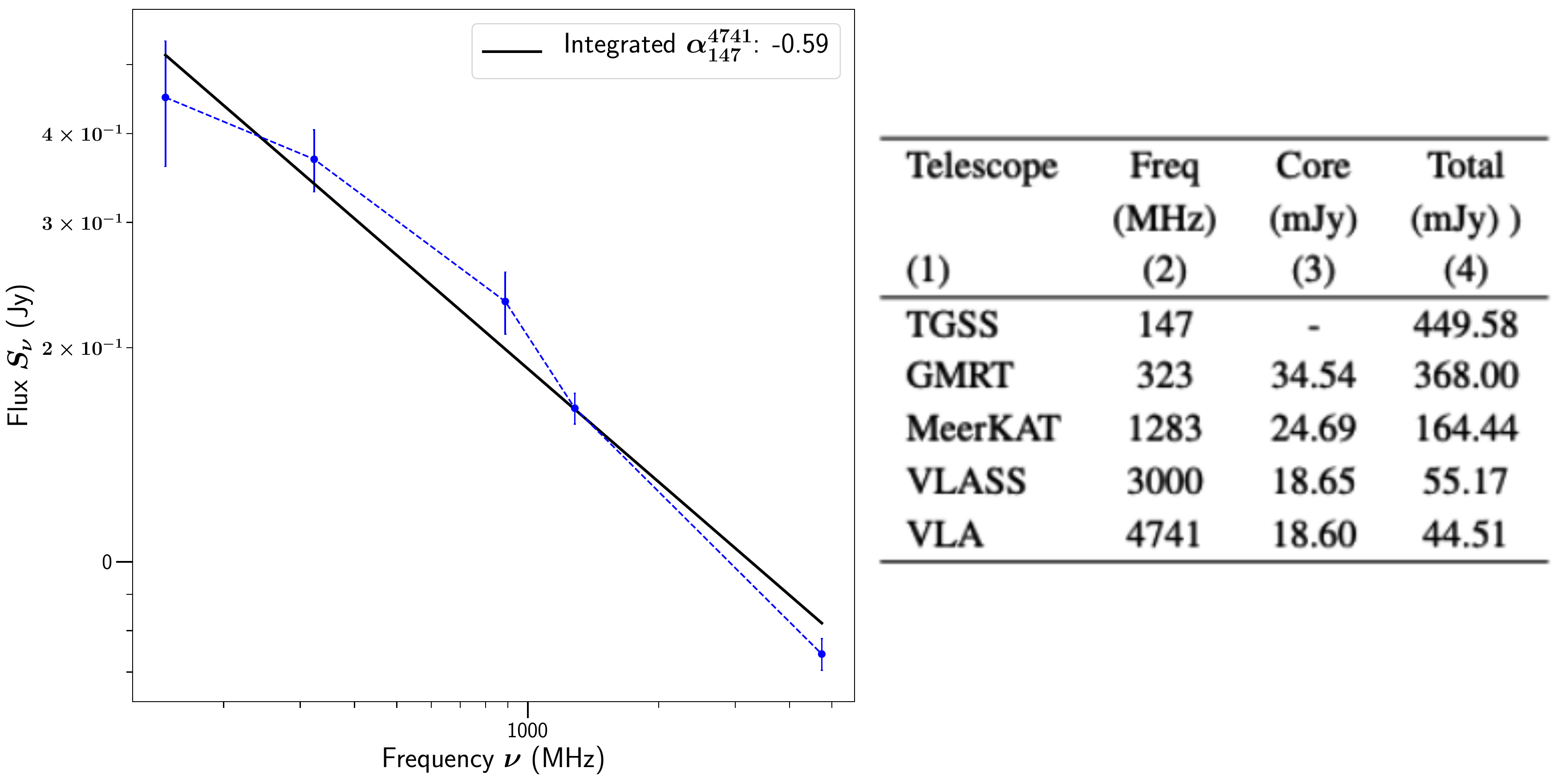}
   \caption{ Spectral fit for integrated flux densities for RAD12 from 147 MHz to 4741 MHz. The table presents flux densities of various components of RAD12 at various radio frequencies.}
   \label{fig:SIFIT}
\end{figure}

In order to understand the overall radio spectral nature of RAD12, we have plotted the integrated flux densities of RAD12 with the available archival data ranging from $\sim$\,150 MHz to 5000 MHz  as seen in Figure.\ \ref{fig:SIFIT}. It is clear that the least square fit to the whole spectrum gives a spectral index of $\alpha$ $\sim$ -0.59. This is similar to the injection spectral index of jets in radio galaxies suggesting either youth or re-acceleration of the old radio plasma (lacking jet or hotspot structure). The fainter outer regions (cap) of the mushroom bubble are steeper compared to the stem or central plume which is similar to FRII radio lobes. The radio bubble extends to nearly 137 kpc and unlike radio bubbles seen in nearby Seyfert galaxies \citep{Hota2006}. The spectral flattening and one-sidedness of the jet is unlikely due to the close alignment of the jet and line of sight as expected for a quasar.  

%\subsection{WISE}
The lack of broad emission lines in the optical spectrum of the host galaxy of RAD12, J004300.63–091346.3, as seen in  Sloan Digital Sky Survey (SDSS), also does not favour the quasar nature of the AGN. Using the Wide-field Infrared Survey Explorer survey \citep[\textit{WISE};][]{WRIGHTWISE} we obtain the mid-infrared (mid-IR) magnitudes of RAD12. 
\textit{WISE} mid-IR colours  can be used to differentiate between quasars, star-forming, high and low excitation AGNs (see figure 3 of \citealt{Dabhade2020}). We adopt the scheme of \citet{Dabhade2020} and references therein to classify RAD12 as a low-excitation radio galaxy based on its colours (W1-W2 = 0.08 and W2-W3=1.19).

%---------comparison
There are nearly half a dozen objects displaying radio jet hitting a neighbouring galaxy. They are Minkowski's Object near NGC541 \citep{Croft2006}, Death Star Galaxy or 3C321 \citep{Evans2008}, 3C34, 3C285, 3C441 and now RAD12 \citep[][; and references therein]{HOTA2022}. 
Hence, it is evident that even a dwarf galaxy can cause morphological distortions in the overall radio structure if it is in the direct path of a radio jet. Interaction with stellar shells has been proposed earlier to explain jet bending in Cen A \citep{GKS84}. Given the prominent stellar shells (Figure.\ \ref{fig:color}b), we propose that the jet in RAD12 has likely been stopped by interstellar medium (ISM) trapped in the stellar shells. A preliminary look at the X-ray emission from the XMM-Newton data hints at the presence of a cavity, which needs further detailed investigation.
From the limited radio-optical data available in archives and 323 MHz follow-up with the GMRT the newly discovered radio source RAD12 is demonstrated, in this very first imaging study, to be unique, justifying deep multi-wavelength follow-up by the community in radio polarisation, X-ray imaging and spectroscopy of ionised/molecular/atomic gas observations. It will be interesting if the gas kinematics can reveal how the radio bubble interacts with the ISM of the merging system, and hence will be very informative for AGN feedback studies.
%%%%%%%%%%%%% 

\vspace{-0.3cm}
\section{Acknowledgement}
We acknowledge financial support from the University Grants Commission (India), and International Astronomical Union for RAD@home team members due to which they could  travel internationally and contribute to oral presentations, train international students for citizen science research and conduct outreach events. RAD@home thanks UM-DAE Centre for Excellence in Basic Sciences (Mumbai), Nehru Planetarium (NMML, Delhi) and International Centre for Theoretical Sciences - Tata Institute of Fundamental Research (Bengaluru (Code: ICTS/RADatICTS2018/05)) for hosting RAD@home Discovery Camps and contributing to the discovery of RAD12.

\end{document}